\documentclass[twocolumn,showpacs,preprintnumbers,amsmath,amssymb,superscriptaddress]{revtex4}
\usepackage{graphicx}
\usepackage{dcolumn}
\usepackage{bm}
\usepackage{longtable}
\usepackage[countmax]{subfloat}
\usepackage{epstopdf}

\providecommand{\be}{\begin{equation}}
  \providecommand{\ee}{\end{equation}}
\providecommand{\bea}{\begin{eqnarray}}
  \providecommand{\eea}{\end{eqnarray}}
\providecommand{\beas}{\begin{eqnarray*}}
  \providecommand{\eeas}{\end{eqnarray*}}

\providecommand{\beni}{\begin{equation*}}
  \providecommand{\eeni}{\end{equation*}}

\providecommand{\bw}{\begin{widetext}}
  \providecommand{\ew}{\end{widetext}}

\usepackage{epsfig}
\usepackage{amssymb}
\usepackage{amsmath}
\usepackage{amsthm}
\usepackage{latexsym}
\usepackage{graphicx}
\def\be{\begin{equation}}
\def\ee{\end{equation}}
\def\bc{\begin{center}}
\def\ec{\end{center}}

\def\be{\begin{equation}}
\def\ee{\end{equation}}
\def\bea{\begin{eqnarray}}
\def\eea{\end{eqnarray}}

\begin{document}

\preprint{APS/123-QED}

\title{From Dyson to Hopfield: Processing on hierarchical networks}

\author{Elena Agliari}
\affiliation{Dipartimento di Fisica, Sapienza Universit\`a di Roma, P.le A. Moro 2, 00185, Roma, Italy.}
\author{Adriano Barra}
\affiliation{Dipartimento di Fisica, Sapienza Universit\`a di Roma, P.le A. Moro 2, 00185, Roma, Italy.}
\author{Andrea Galluzzi}
\affiliation{Dipartimento di Matematica, Sapienza Universit\`a di Roma, P.le Aldo Moro 2, 00185, Roma, Italy.}
\author{Francesco Guerra}
\affiliation{Dipartimento di Fisica, Sapienza Universit\`a di Roma, P.le A. Moro 2, 00185, Roma, Italy.}
\author{Daniele Tantari}
\affiliation{Dipartimento di Matematica, Sapienza Universit\`a di Roma, P.le Aldo Moro 2, 00185, Roma, Italy.}
\author{Flavia Tavani}
\affiliation{Dipartimento SBAI (Ingegneria), Sapienza Universit\`a di Roma, Via  A.  Scarpa 14, 00185, Roma, Italy.}

\date{\today}

\begin{abstract}
We consider statistical-mechanics models for spin systems built on hierarchical structures, which provide a simple example of non-mean-field framework. We show that the coupling decay with spin distance can give rise to peculiar features and phase diagrams much richer that their mean-field counterpart.
In particular, we consider the Dyson model, mimicking ferromagnetism in lattices, and we prove the existence of a number of meta-stabilities, beyond the ordered state, which get stable in the thermodynamic limit. Such a feature is retained when the hierarchical structure is coupled with the Hebb rule for learning, hence mimicking the modular architecture of neurons, and gives rise to an associative network able to perform both as a serial processor as well as a parallel processor, depending crucially on the external stimuli and on the rate of interaction decay with distance; however, those emergent multitasking features reduce the network capacity with respect to the mean-field counterpart. The analysis is accomplished through statistical mechanics, graph theory, signal-to-noise technique and numerical simulations in full consistency. Our results shed light on the biological complexity shown by real networks, and suggest future directions for understanding more realistic models.
\end{abstract}

 \maketitle

In the last decade, extensive research on complexity in networks has evidenced (among many results \cite{newman,vespignani}) the widespread of modular structures and the importance of quasi-independent communities in many research areas such as neuroscience \cite{bullmore, kumar}, biochemistry \cite{gallos} and genetics \cite{conaco}, just to cite a few. In particular, the modular, hierarchical architecture of cortical neural networks has nowadays been analyzed in depths \cite{moretti}, yet the beauty revealed by this investigation is not captured by the statistical mechanics of neural networks, nor standard ones (i.e. performing serial processing) \cite{hopfield,amit} neither multitasking ones (i.e. performing parallel processing) \cite{prlnoi,peter}. In fact, these models are intrinsically mean-field, thus lacking a proper definition of metric distance among neurons.

Hierarchical structures have been proposed in the past as (relatively) simple models for ferromagnetic transitions beyond the mean-field scenario -the Dyson hierarchical model (DHM) \cite{dyson}- and are currently experiencing a renewal interest for understanding glass transitions in finite dimension \cite{REM,DH}.
Therefore, times are finally ripe for approaching neural networks embedded in a non-mean-field architecture, and this letter summarizes our findings on associative neural networks where the Hebbian kernel is coupled with the Dyson topology.
%

First, we start studying the DHM mixing the Amit-Gutfreund-Sompolinsky ansatz approach \cite{amit} (to select candidable retrievable states) with the interpolation technique (to check their thermodynamic stability) and we show that, as soon as ergodicity is broken, beyond the ferromagnetic/pure state (largely discussed in the past, see e.g., \cite{Gallavotti,Sinai}), a number of metastable states suddenly appear and become {\em stable} in the thermodynamic limit. The emergence of such states implies the breakdown of classical (mean-field) self-averaging and stems from the weak ties connecting distant neurons, which, in the thermodynamic limit, effectively get split into detached communities (see Fig.~$1$).  As a result, if the latter are initialized with opposite magnetizations, they remain stable.

This is a crucial point because, once implemented the Hebbian prescription to account for multiple pattern storage, it allows proving that the system not only executes extensive serial processing \`a la Hopfield, but its communities perform autonomously, hence making parallel retrieval feasible too. We stress that this feature is essentially due to the notion of metric the system is endowed with, differently from the parallel retrieval performed by the mean-field multitasking networks which require blank pattern entries \cite{prlnoi,peter}.

\begin{figure}[tb] \begin{center}
\includegraphics[width=.22\textwidth]{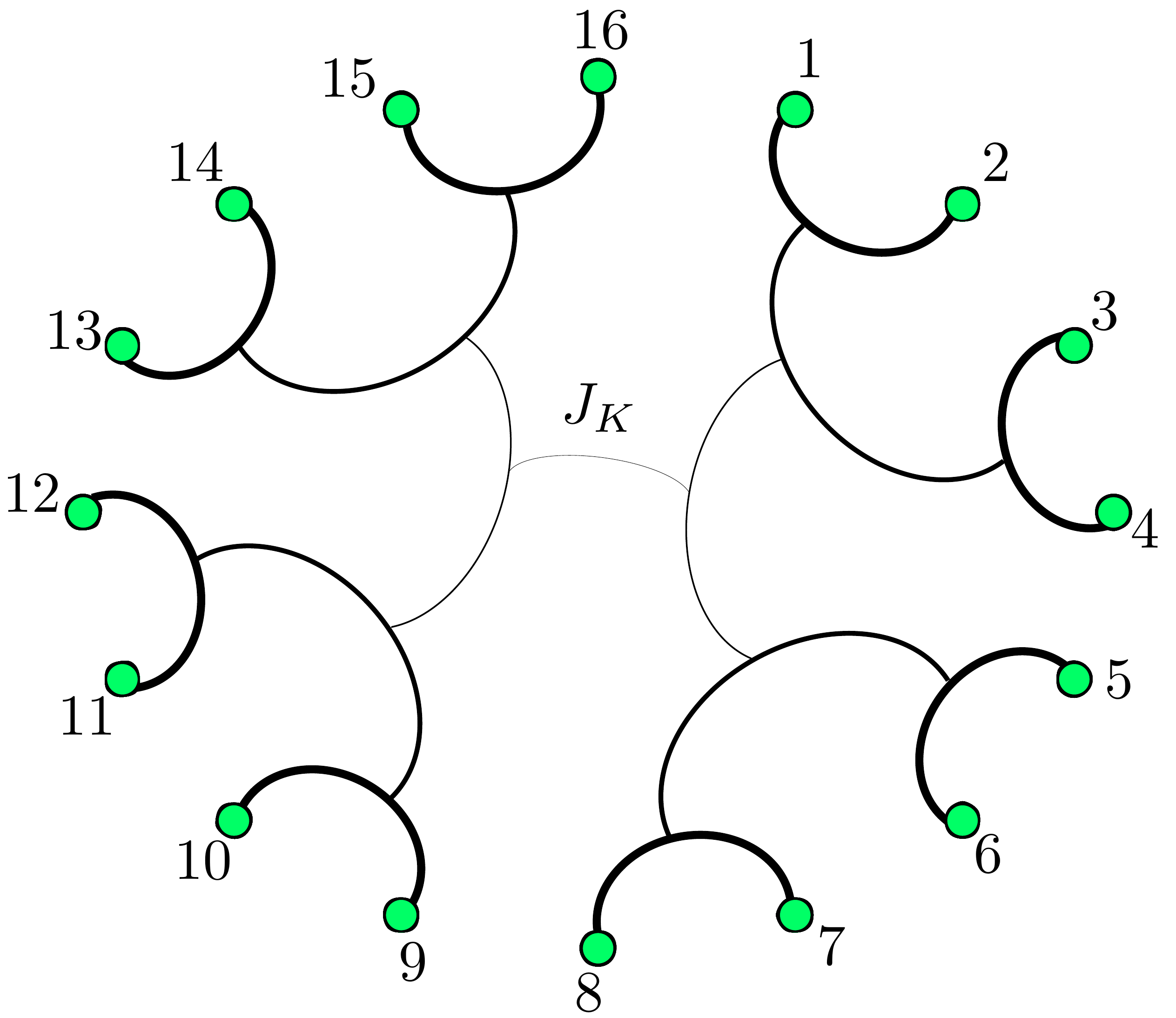}
\includegraphics[width=.22\textwidth]{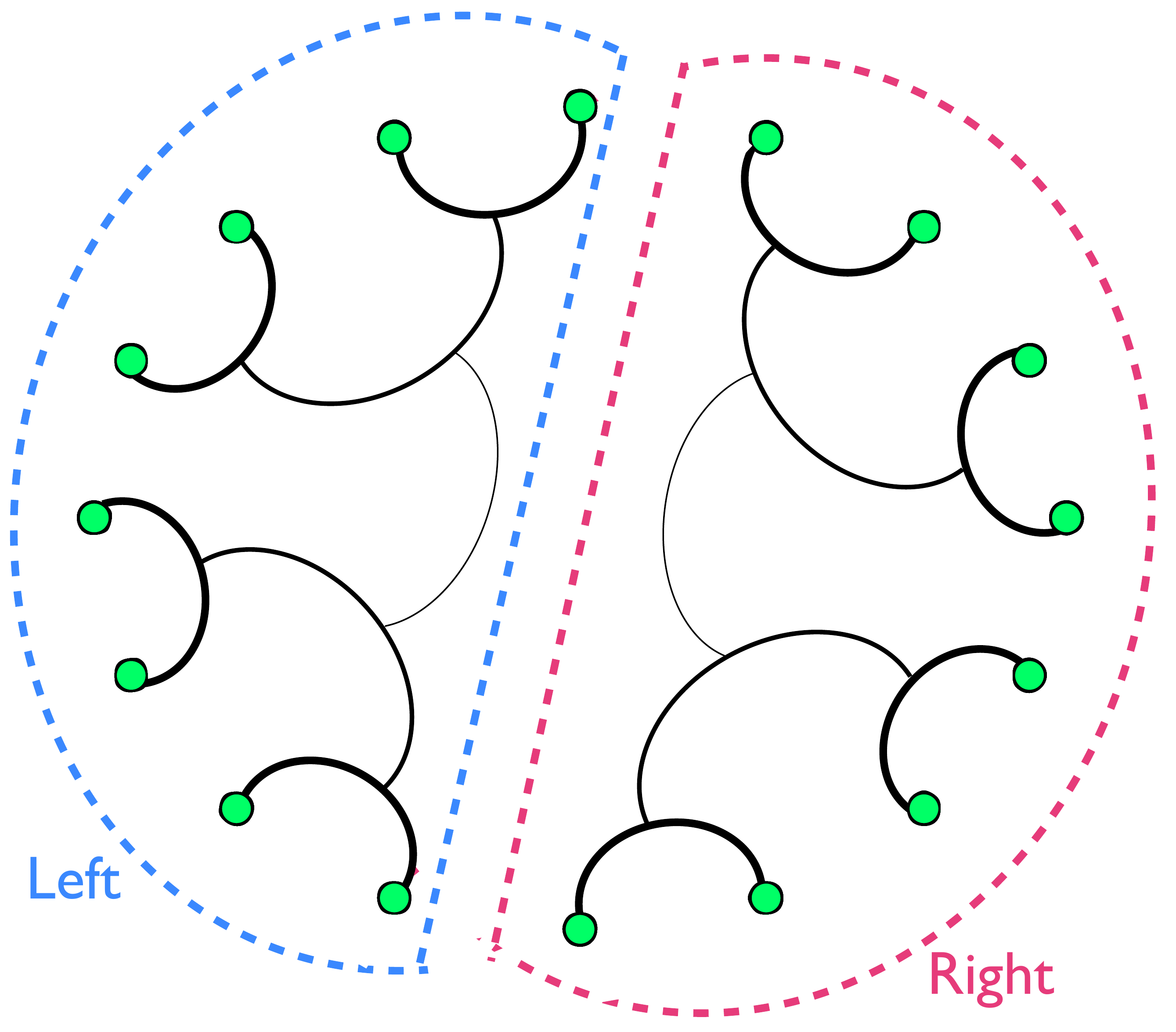}
\caption{\label{fig:esempio} Schematic representation of the hierarchical topology where the associative network insists. Green spots represent Ising neurons ($N=16$ in this shapshot). The larger the distance among spins the weaker their coupling (see eq. \ref{zozzi}).
}
\end{center}
\end{figure}

Therefore, the hierarchical neural network is able to perform both as a serial processor and as a parallel processor.
We corroborate this scenario merging results from statistical mechanics, graph-theory, signal-to-noise technique and extensive numerical simulations as explained hereafter.

In the DHM the mutual interaction between $2^{k+1}$ Ising spins $\sigma_i=\pm 1$, with $i=1,...,2^{k+1}$, is described by  the following Hamiltonian defined recursively as
\be\label{disonne}\small\small\small
H_{k+1}(\vec{\sigma})=H_k(\vec{\sigma_1})+H_k(\vec{\sigma_2})-\frac{J}{2^{2\rho(k+1)}}\sum_{i<j=1}^{2^{k+1}}\sigma_i\sigma_j,
\ee
where $J>0$ and $\rho \in ]1/2,1[$ tune the interaction strength, $\vec{\sigma_1}\equiv \{ \sigma_i \}_{1 \leq i \leq 2^k}$,  $\vec{\sigma_2}\equiv \{ \sigma_j \}_{2^k+1 \leq j \leq 2^{k+1}}$ and $H_0(\vec{\sigma})=0$.
\newline
This model is explicitly non-mean-field as we implicitly introduced a distance: Two spins $i$ and $j$ turn out to be at distance $d_{ij}=d$ if, along the recursive construction, they first get connected at the $d$-th  iteration; of course $d$ ranges in $[1,k]$ (see also Fig.~$1$).
It is possible to re-write the Hamiltonian (\ref{disonne}) straightforwardly in terms of $d_{ij}$ as $H_{k+1}(\vec{\sigma})= -\sum_{i<j}J_{ij}\sigma_i \sigma_j$, being
\be\small\label{zozzi}
J_{ij}=\sum_{l=d_{ij}}^{k} \left ( \frac{J}{2^{2\rho l}} \right)=J(d_{ij},k,\rho)=J \frac{4^{\rho -d_{ij}  \rho }-4^{-(k+1) \rho }}{4^{\rho }-1}.
\ee
Set the noise level $\beta=1/T$ in proper units, we are interested in an explicit expression of the infinite volume limit of the mathematical pressure $\alpha(\beta,J,\rho)=-\beta f(\beta,J,\rho)$, (where $f$ is the free energy)  defined as
\be\small
\nonumber
\alpha(\beta,J,\rho)= \lim_{k\to\infty} \frac{1}{2^{k+1}}\log \sum_{\vec{\sigma}}\exp [ -\beta H_{k+1}(\vec{\sigma})+ h \sum_{i=1}^{2^{k+1}}\sigma_i],
\ee
whose maxima correspond to equilibrium states. In particular, we want to find such extremal points with respect to the global magnetization $ m_{k+1}=\frac{1}{2^{k+1}}\sum_i^{2^{k+1}}\sigma_i$ and the set of $k$ magnetizations $\vec{m}_1,..., \vec{m}_k$, which quantify the state of each community, level by level; the two magnetizations related to the two largest communities (see Fig.$1$) read off as
\be\small
\nonumber
m^{(1)}_{k} = m_{\textrm{left}}=\frac{1}{2^{k}}\sum_{i=1}^{2^{k}}\sigma_i, \ \ \ m^{(2)}_{k} = m_{\textrm{right}}=\frac{1}{2^{k}}\sum_{i=2^k+1}^{2^{k+1}}\sigma_i.
\ee
We approach the investigation of the DHM meta-stabilities exploiting the interpolative technology introduced in \cite{DH}, that allows obtaining bounds beyond the mean-field paradigm (as fluctuations are not completely discarded). This procedure returns  the following expression for the pure ferromagnetic (i.e., $m_{\textrm{left}}=m_{\textrm{right}}=m$) pressure (see \cite{DH,nostro} for details)
\be\small\label{NMFbound}
\alpha(\beta,J,\rho) \geq  \sup_{m} \Big \{ \log 2+\log\cosh [ \beta (h +   J m C_{2\rho})]- \frac{\beta J m^2}{2} C_{2\rho} \Big \},
\ee
%
where $C_y=2^y/[(2^{y}-1)(2^y-2)]$. However, let us suppose that the two main communities (left and right) display different magnetizations $m_{\textrm{left}}=m_1$ and $m_{\textrm{right}}=m_2$: formula $(\ref{NMFbound})$, implicity derived within the ansatz of pure ferromagnetic state, can therefore be generalized within the ansatz of mixed state (i.e., $m_{\textrm{left}} = - m_{\textrm{right}}$) as
\begin{eqnarray}\small \nonumber
\alpha(\beta,J,\rho) &\geq&  \sup_{m_1, m_2}\Big\{\ln 2 -\frac{\beta J}{2} C_{2\rho} \Big (\frac{m_1^2+m_2^2}{2} \Big)\\ \small
&+&\frac{1}{2} [L(\beta m_1 C_{2\rho}) \nonumber
+L(\beta m_2 C_{2\rho} )] \Big \},
\end{eqnarray}
where $L(x)=\ln\cosh(x)$ (of course, posing $m_1=m_2=m$, we recover the former bound).
Requiring thermodynamic stability we obtain the following self-consistencies
\be \label{eq:SC1}
m_{1,2}=\tanh[h+\beta Jm_{1,2} C_{2\rho}],
\ee
whose solution is successfully compared with data from Monte Carlo (MC) simulations in Fig.~$3$.

As can be derived from Eq.~\ref{disonne}, the meta-stable mixed/parallel state and the stable ferromagnetic/serial state display an (intensive) energy gap $\Delta E \propto 1/2^{(k+1)(2\rho-1)}$ hence, while thermodynamics is dominated by the ferromagnetic/serial behavior, for $k\to\infty$ both the states become stable (see Fig.~$2$) sharing the same intensive free-energy. This can be easily confirmed by the stability analysis, as the Hessian of $\alpha(\beta,J,\rho)$ actually depends on $m_1^2$ and on $m_2^2$ only \cite{nostro} hence, as the paramagnetic solution becomes unstable, both the ferromagnetic/serial (i.e. $m_{\textrm{left}}=m_{\textrm{right}}$) and the  mixed/parallel, (i.e.  $m_{\textrm{left}}= -m_{\textrm{right}}$) solutions appear.
\begin{figure}\label{fig:cluster}
\includegraphics[width=8.5cm]{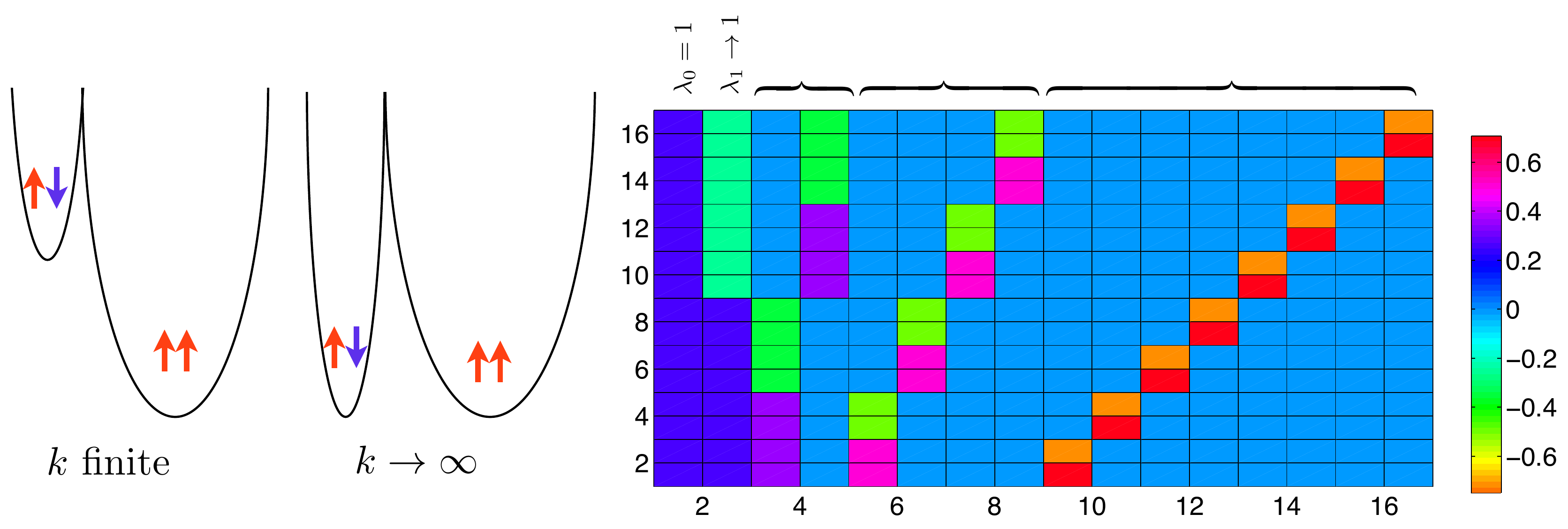}
\caption{Left panel: Sketch of ferromagnetic and mixed free energy minima for the DHM at finite size and in the thermodynamic limit. Right panel: Representation of the eigenstates of $T$ for a system with $k=6$ and $\rho=0.75$. Each column represents a different eigenstate, eigenstates pertaining to the same degenerate eigenvalue are highlighted. Different colors represent different entries in the eigenstate, as shown by the colormap on the right.}
\end{figure}
This point can be further understood by a graph-theoretical approach. The DHM can be looked at as a ferromagnet embedded in a fully-connected topology, where the link connecting two arbitrary nodes $i$ and $j$, displays a weight $J_{ij}$ decaying with the distance between $i$ and $j$, and defined according to a suitable metric (e.g., the one based on recursion described above or the $2$-adic metric $\tilde{d}_{ij} = 2^{-\textrm{ord}_2(i-j)}$, in such a way that $J_{ij} \sim \tilde{d}_{ij}^{-2 \rho}$ \cite{nostro}). This structure exhibits a high degree of modularity and of clustering \cite{nostro}. Moreover, the set of nodes is countable and weights are finite, i.e. $J_{\textrm{min}}=4^{-(k+1) \rho} \leq J_{ij} \leq J_{\textrm{max}}=(1 - 4^{-(k+1) \rho})/(4^{\rho} -1)$, thus, upon proper normalization of weights $J_{ij} \rightarrow T_{ij} = J_{ij} / w_i$, where $w_i = \sum_j J_{ij}$, the graph describes a Markov chain, where each node represents a state and $T$ is the transition matrix \cite{Norris}.  The evolution of the random process is therefore provided by the master equation $p(t+1) = T p(t) \rightarrow \dot{p}(t) = T p(t) - p(t)$, whose stationary distribution, referred to as $\pi$, satisfies $\pi = T \pi$, that is, $\pi$ coincides with the eigenvector $\phi_{\lambda_0}$ of $T$ corresponding to eigenvalue $\lambda_0=1$ (that is just the Perron-Frobenius eigenvalue of $T$) and it is uniformly distributed as $\pi= \mathbf{e}/2^{(k+1)/2}$.
\begin{figure}\label{fig:W_Hop}
\includegraphics[width=8.5cm]{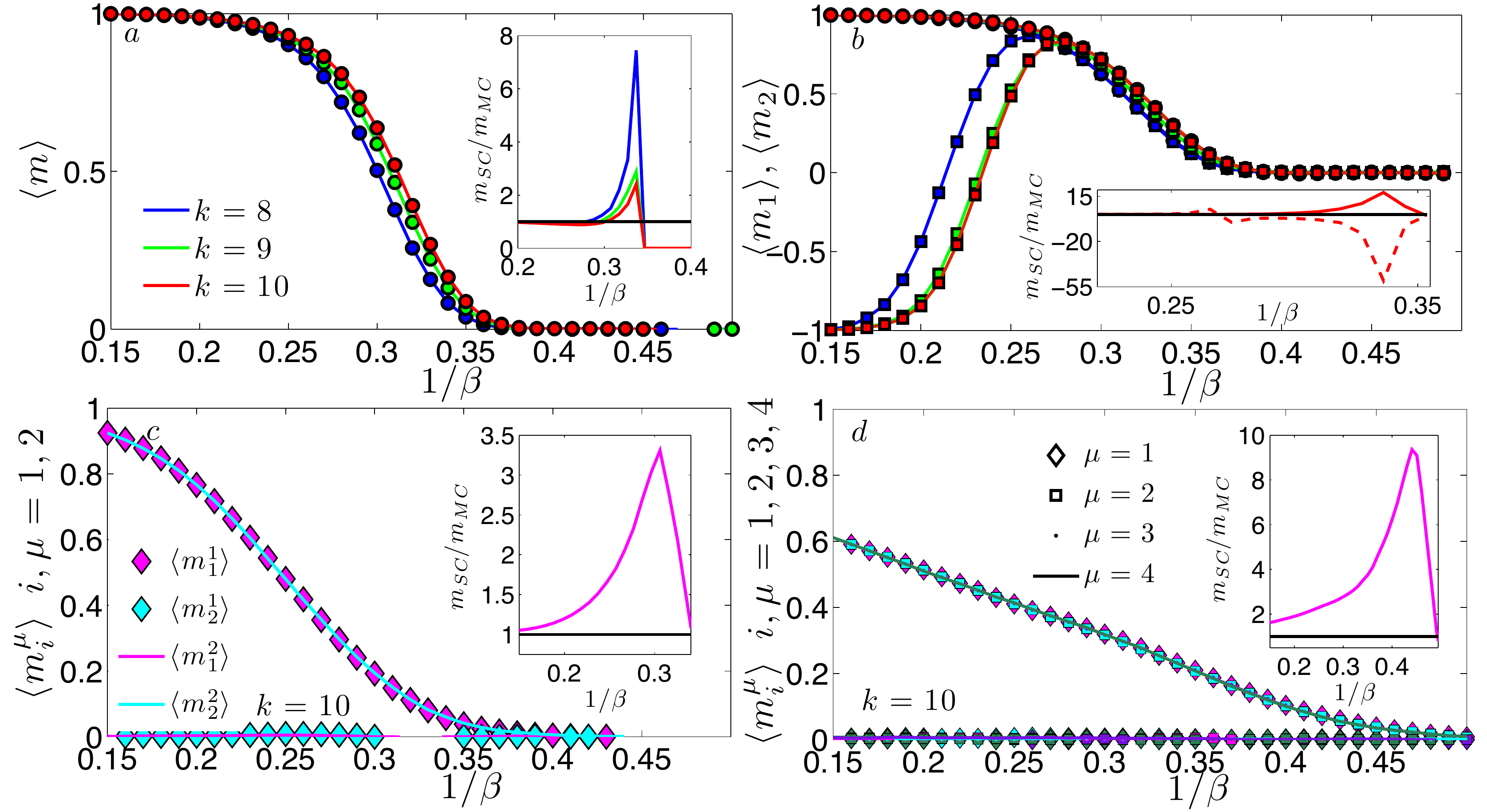}
\caption{Panels $a$ and $b$: Magnetizations obtained via MC simulations of the DHM for different sizes (main figure) and comparison with theoretical curves given by Eq.~$4$ (insets). Notice that the spontaneous switch between serial and parallel state in panel $b$ is a finite-size effect. Lower panels: Mattis magnetizations obtained via MC simulations of the HHM (main figures) and comparison with theoretical curves given by Eq.~\ref{eq:SC2} (insets) for $p=2$ (panel $c$) and for $p=4$ (panel $d$). The noise level in analytical results was rescaled to collapse $\beta_c^{-1}$ with the one numerically estimated via Binder cumulants.}
\end{figure}
Moreover, the second-largest eigenvalue $\lambda_1$, and the related eigenstate are, respectively
\begin{eqnarray}\small \nonumber
\lambda_1 &=& \sum_{j=1}^{2^{k}} T_{1j} - 2^{k} T_{12^{k+1}} \rightarrow 1 - \mathcal{ O}(2^{-(2\rho-1)(k+1)}),\\
\nonumber
\phi_{\lambda_1} &=& (\underbrace{1, 1, \dots, 1}_{2^{(k+1)/2}}, \underbrace{ -1, -1, \dots, -1}_{2^{(k+1)/2}})/2^{(k+1)/2}.
\end{eqnarray}
As $\lambda_1$ converges to $1$ in the thermodynamic limit, ergodicity breaking for the stochastic process is expected. In fact, $\phi_{\lambda_0}$ and $\phi_{\lambda_1}$ generate a subspace where any vector is an eigenvector of $T$ with the same eigenvalue $\lambda = 1$. In particular, we see that
\bea\small
\phi_{\lambda_0} + \phi_{\lambda_1} = (\underbrace{1, 1, \dots, 1}_{2^{(k+1)/2}}, \underbrace{ 0, 0, \dots, 0}_{2^{(k+1)/2}}) \sqrt{2/2^{k+1}},\\ \small
\phi_{\lambda_0} - \phi_{\lambda_1} = (\underbrace{0, 0, \dots, 0}_{2^{(k+1)/2}}, \underbrace{ 1, 1, \dots, 1}_{2^{(k+1)/2}}) \sqrt{2/2^{k+1}},
\eea
correspond to stationary states localized on the left and on the right branch of the graph, respectively. Otherwise stated, there is no flow between the two main branches as if they were autonomous. The same holds as we split each branch in smaller sub-units iteratively (see Fig.~$2$), and mirrors the genesis of metastable states in the thermodynamics side.

As a final perspective, we check the robustness of states through a signal-to-noise analysis. To this aim we explicit the fields insisting on the spins in (\ref{disonne}) by writing $H_{k+1}(\vec{\sigma}) =\sum_i h_i(\vec{\sigma}|\rho) \sigma_i$, being
\be\small\
h_i(\vec{\sigma} |\rho)= \sum_{\mu=1}^{k+1} \Big[  \sum_{l=\mu}^{k+1}\Big(\frac{J}{2^{2 l \rho}}\Big) 2^{\mu-1}m_{f(\mu,i)}^{\mu-1}  \Big],
\ee
where $m^{\mu-1}_{f(\mu,i)}$ is the normalized magnetization of spins at distance $\mu$ from the $i$-th one.
The microscopic law governing the evolution of the system  is  a stochastic alignment with the local field $h_i(\vec{\sigma}|\rho)$, that is, $\sigma_i(t+\delta t)=\textrm{sign}\{\tanh[\beta h_i(\mathbf{\vec{\sigma}}(t)|\rho ) ]+\eta_i(t) \}$. In the noiseless limit, the stochasticity captured by the independent random numbers $\eta_i(t)$ (uniformly distributed over the interval $[-1,1]$) is lost, and
\be\nonumber
\lim_{\beta \to \infty} \sigma_i(t+\delta t)=\textrm{sign} \{ h_i(\vec{\sigma}(t)|\rho ) \}.
\ee
Thus, if $\sigma_i h_i(\vec{\sigma}|\rho) >0, \forall i  \in  [1,2^{k+1}]$, the configuration $\vec{\sigma}$ is dynamically stable. Hereafter, we focus on the ferromagnetic/serial case and on the mixed/parallel case only, referring again to \cite{nostro} for an extensive treatment.

In the former case, $\sigma_i=+1,  \forall i \in[1,2^{k+1}]  \Rightarrow   h_i(\vec{\sigma}|\rho)>0 \ \forall k,\rho\in]0.5,1]$. Therefore, the ferromagnetic/serial case state is stable for $\beta \to \infty$ and $ \rho \in]0.5,1]$.

In the latter case, $\sigma_i=+1, \forall i \in[1,2^{k}]$ and $\sigma_i=-1, \forall i \in[2^{k}+1,2^{k+1}] \Rightarrow \lim_{k\to\infty}h_i(\vec{\sigma}|\rho ) = 1 /(2^{1-2 \rho }+4^{\rho }-3)$. Therefore, the mixed/parallel case is stable for $\beta \to \infty$ and $\rho\in ]0.5,1]$.

Clearly, we can iterate this scheme, splitting the largest communities in two, up to $\mathcal{O}(k)$ times.


Now, retaining the outlined perspective, we recursively define the hierarchical Hopfield model (HHM)  by the following Hamiltonian
\be\small
H_{k+1}(\vec{\sigma})=H_k(\vec{\sigma_1})+H_k(\vec{\sigma_2})-\frac{1}{2}\frac{1}{2^{2\rho(k+1)}}\sum_{\mu=1}^{p}\sum_{i,j=1}^{2^{k+1}}\xi^{\mu}_i\xi^{\mu}_j\sigma_i\sigma_j
\ee
with $H_0(\vec{\sigma})=0$, $\rho \in ]1/2,1[$ and where, beyond $2^{k+1}$ dichotomic neurons, also $p$ quenched patterns $\bold{\xi}^{\mu}$, $\mu \in (1,...,p)$ are introduced. Their entries $\xi_i^{\mu}=\pm 1$ are drawn with the same probability $1/2$ and are averaged by $\mathbb{E}_{\xi}$.
\newline
Again, we can write the Hamiltonian of the HHM in terms of the distance $d_{ij}$, obtaining $H_{k+1}(\vec{\sigma})= -\sum_{i<j}\widetilde{J_{ij}}\sigma_i \sigma_j $, where
\be
\widetilde{J_{ij}}=\frac{4^{\rho -d_{ij}  \rho }-4^{-k \rho }}{4^{\rho }-1} \\ \sum_{\mu=1}^p \xi_i^\mu \xi_j^\mu,
\ee
hence the Hebbian kernel on a hierarchical topology is tuned by the distance-dependent weight $J(d_{ij},k,\rho)$.

Once introduced suitably Mattis overlaps, both global $m_{\mu} = \sum_{i=1}^{2^{k+1}}\xi_i^{\mu} \sigma_i/2^{k+1}$, and community restricted, as
\be\small
m^{\mu}_{\textrm{left}}=\frac{1}{2^k}\sum_{i=1}^{2^k}\xi_i^{\mu} \sigma_i, \ %
m^{\mu}_{\textrm{right}}=\frac{1}{2^k}\sum_{j=2^k+1}^{2^{k+1}}\xi_j^{\mu} \sigma_j,
\ee
the statistical-mechanical route returns a non-mean field approximation for the pressure of the serial-retrieval state as
$\alpha \geq \sup_{m} \{ \log 2 -\frac{\beta}{2}\sum_{\mu=1}^p m_{\mu}^2 C_{2\rho}
+\mathbb{E}_{\xi}\log\cosh [ \sum_{\mu=1}^{p}(\beta m_{\mu} C_{2\rho})\xi^{\mu} ] \}$,
with optimal order parameters fulfilling
\be
m^{\mu}=\mathbb{E}_{\xi}\xi^{\mu}\tanh \{ \beta\sum_{\nu=1}^p ( C_{2\rho} m^{\nu} ) \xi^{\nu} \},
\ee
and critical temperature $\beta_c^{-1}=C_{2\rho}$.
Assuming two different families of Mattis magnetizations $\{m^{\mu}_{1,2}\}_{\mu=1}^p$ for the largest communities (left and right), we get a non-mean-field approximation for parallel-retrieval pressure
\begin{eqnarray}\small
\nonumber
\alpha &\geq& \sup_{\{m^{\mu}_{1,2}\}} \{ \ln 2-\frac{\beta}{2} C_{2\rho} \sum_{\mu=1}^p \frac{(m^{\mu}_1)^2+(m^{\mu}_2)^2}{2}\\
\nonumber
&+&\frac12\mathbb{E}_{\xi} [ L(\sum_{\mu=1}^{p}(\beta m^{\mu}_1 C_{2\rho} +\xi^{\mu})) + L(\sum_{\mu=1}^{p}(\beta m^{\mu}_2 C_{2\rho}  )) ] \} ,
\end{eqnarray}
whose disentangled optimal order parameters satisfy
\be \label{eq:SC2}
m^{\mu}_{1,2}=\mathbb{E}_{\xi} \{ \xi^{\mu}\tanh [ \beta\sum_{\nu=1}^p C_{2\rho} m^{\nu}_{1,2}\xi^{\nu}] \},
\ee
returning again $\beta_c^{-1}=C_{2\rho}$ and the behavior sketched in Fig.$3$. We could use this argument iteratively splitting the system in smaller and smaller blocks: in $M$ times, we have to use different magnetizations (for these $2^M$ small communities) until the $k+1-M$ level. The procedure keeps working as far as $\lim_{k\to\infty} \sum_{l=k-M}^k 2^{l(1-2\rho)} \sum_{\mu=1}^p m^{\mu}_{l}=0$, hence, if we want the system to cope with $p$ patterns at once, we need $p$ different blocks, thus $M=\log(p)$ and at best, as $2^{(1-2\rho)[k-\log(p)]}$, $p/2^k \to 0$ as $k\to\infty$, $p \leq \mathcal{O}(k)$.

\begin{figure}[tb]\label{fig:segnali}
\includegraphics[width=8.3cm]{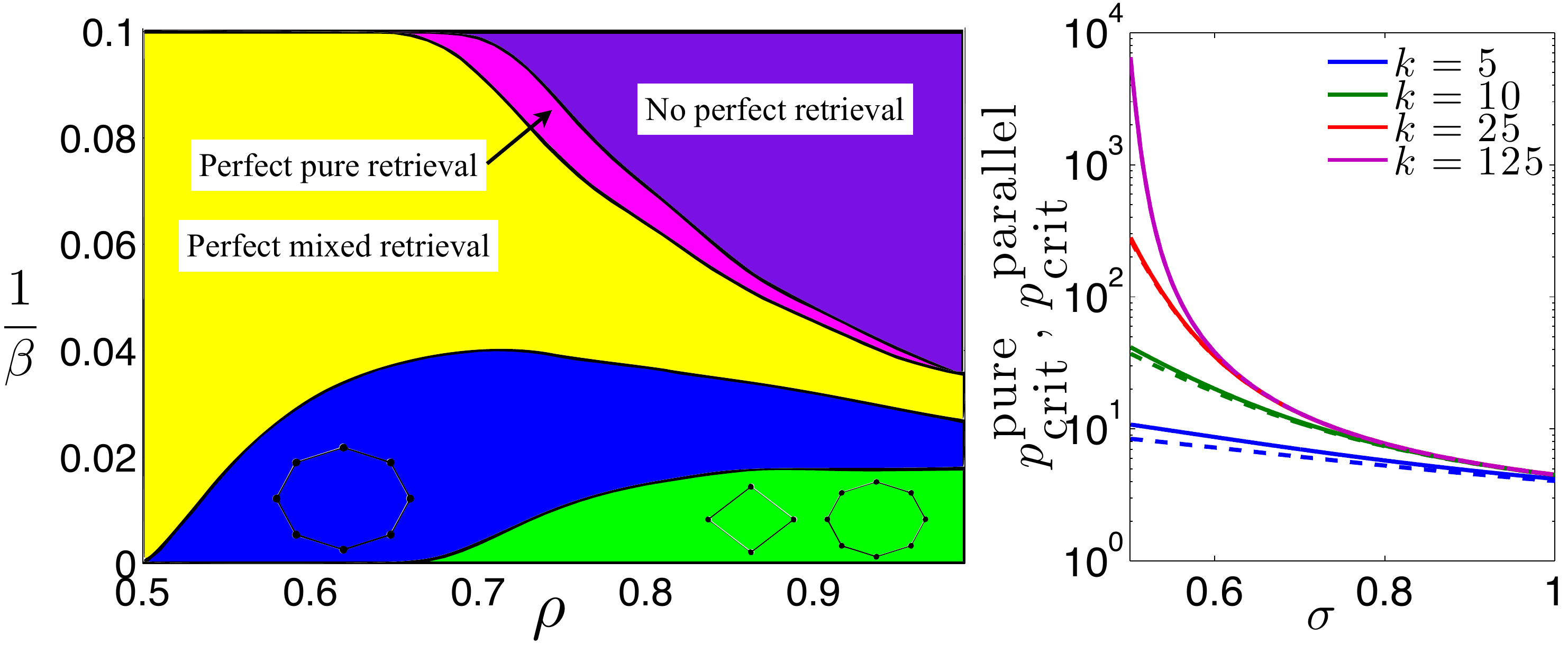}
\caption{Left panel: Phase diagram for the DHM as derived from the signal-to-noise analysis. The curves separating different phases are obtained by solving numerically the transcendental equation $\tanh [\beta h_i( \vec{\sigma}, \rho, k) ]=1$ as a function of $\beta$ and $\rho$. Here we fixed $k=7$ and we focused on four different configurations (pure state, parallel state and states where sub-communities made of four and eight spins, respectively, are misaligned with respect to the bulk). Right panel: $p_{\textrm{crit}}^{\textrm{pure}}$ (solid line) and $p_{\textrm{crit}}^{\textrm{parallel}}$ (dashed line), as a function of $\sigma$ and for several choices of $k$, as explained by the legend.
}
\end{figure}

This picture is confirmed by the signal-to-noise analysis: we start from the pure state, i.e., $\sigma_i=\xi_i^{\mu}$, and check its stability writing $\sigma_i h_i (\vec{\sigma}|\rho)$  as a signal term plus a noise term and then comparing their amplitudes:
\be\small
\xi_i^\mu h_i (\vec{\sigma}|\rho)=\xi_i^\mu \sum_{\nu=1}^p \xi_i^\nu \sum_{d=1}^k J(d,K,\rho)\sum_{j: d_{ij}=d}\xi_j^\nu \xi_j^\mu = S+ R(\xi),
\ee
where $S=\sum_{d=1}^k J(d,k,\rho) 2^{d-1}\geq 0$, while
$$
R(\xi)=\xi_i^\mu \sum_{\nu\neq\mu}^p \xi_i^\nu \sum_{d=1}^k J(d,k,\rho)\sum_{j: d_{ij}=d}\xi_j^\nu \xi_j^\mu.
$$
As clearly $\langle R(\xi) \rangle_\xi=0$, we need to evaluate when the ratio $S/|\sqrt{\langle R(\xi)^2 \rangle_\xi}\to 1$: the latter returns, the maximum load $p_{\textrm{crit}}^{\textrm{pure}}(k,\rho)$ storable by the network before the noise prevails over the signal and retrieval becomes forbidden.
\newline
As for parallel-retrieval stability, forcing $\sigma_i=\xi_i^\mu $ $\forall i \in[1,2^{k}]$ and $\sigma_i=\xi_i^\gamma $ $\forall i \in[2^{k}+1,2^{k+1}]$ for $\mu\neq\gamma$, and splitting again $\sigma_i h_i(\vec{\sigma}|\rho)$ in a signal plus a noise term, we can check again the maximum load $p_{\textrm{crit}}^{\textrm{parallel}}(k,\rho)$ storable by the network. Both $p_{\textrm{crit}}^{\textrm{pure}}(k,\rho)$ and $p_{\textrm{crit}}^{\textrm{parallel}}(k,\rho)$ are monotonically decreasing functions of $\rho$, and they converge to the finite value $(4^{\rho}-1)(4^{2\rho} -2)/(-3\times 4^{\rho} +4^{2\rho} +2)^2+1$ as $k$ gets larger (see Fig.~$4$ and \cite{nostro} for more details).

Summarizing, beyond classical retrieval, the network is able to safely handle multiple patterns, too. However, there is a cost in terms of capacity: as we can neglect weak links among upper levels - those that become effectively negligible in the thermodynamic limit -  communities perform autonomously, yet, globally, the network loses a significant amount of bit-storing synapses. Thus a new compromise appear in non-mean-field cognitive systems: increasing multitasking capabilities diminishes the processor capacity, the trigger between them being ruled mainly by the rate of interaction decay $\rho$.

\vspace{0.15cm}

Sapienza University, GNFM-INdAM and INFN are acknowledged for financial support.

\end{document}